\documentstyle[12pt]{article}

 \catcode`\@=11 

\def\maketitle{\par
 \begingroup
 \def\thefootnote{\fnsymbol{footnote}}
 \def\@makefnmark{\mbox{$^\@thefnmark$}}
 \@maketitle
 \@thanks
 \endgroup
 \setcounter{footnote}{0}
 \let\maketitle\relax
 \let\@maketitle\relax
 \gdef\@thanks{}\gdef\@author{}\gdef\@title{}\let\thanks\relax}
\def\@maketitle{\vspace*{0.9cm}
{\hsize\textwidth
 \linewidth\hsize \centering
 {\normalsize \bf \@title \par} \vskip 1.0cm  {\normalsize  \@author \par}}}

\def\thefootnote{\mbox{\noindent$\fnsymbol{footnote}$}}
    \long\def\@makefntext#1{\noindent$^{\@thefnmark}$#1}

\def\section{\@startsection{section}{1}{\z@}{1.5ex plus 0.5ex minus
   1.2ex}{1.3ex plus .1ex}{\normalsize\bf}}
\def\subsection{\@startsection{subsection}{2}{\z@}{1.5ex plus 0.5ex minus
    1.2ex}{1.3ex plus .1ex}{\normalsize\em}}

\def\@sect#1#2#3#4#5#6[#7]#8{\ifnum #2>\c@secnumdepth
     \def\@svsec{}\else
     \refstepcounter{#1}\edef\@svsec{\ifnum #2=1 \@sectname\fi
        \csname the#1\endcsname.\hskip 1em }\fi
     \@tempskipa #5\relax
      \ifdim \@tempskipa>\z@
        \begingroup #6\relax
          \@hangfrom{\hskip #3\relax\@svsec}{\interlinepenalty \@M #8\par}
        \endgroup
       \csname #1mark\endcsname{#7}\addcontentsline
         {toc}{#1}{\ifnum #2>\c@secnumdepth \else
                      \protect\numberline{\csname the#1\endcsname}\fi
                    #7}\else
        \def\@svsechd{#6\hskip #3\@svsec #8\csname #1mark\endcsname
                      {#7}\addcontentsline
                           {toc}{#1}{\ifnum #2>\c@secnumdepth \else
                             \protect\numberline{\csname the#1\endcsname}\fi
                       #7}}\fi
     \@xsect{#5}}

\def\@sectname{}

\def\thebibliography#1{\section*{{{\normalsize
\bf References }
\rule{0pt}{0pt}}\@mkboth
  {REFERENCES}{REFERENCES}}\list
  {{\arabic{enumi}.}}{\settowidth\labelwidth{{#1}}%
    \leftmargin\labelwidth  \frenchspacing
    \advance\leftmargin\labelsep
    \itemsep=-0.2cm
    \usecounter{enumi}}
    \def\newblock{\hskip .11em plus .33em minus -.07em}
    \sloppy
    \sfcode`\.=1000\relax}

\def\@cite#1#2{\unskip\nobreak\relax
    \def\@tempa{$\m@th^{\hbox{\the\scriptfont0 #1}}$}%
    \futurelet\@tempc\@citexx}
\def\@citexx{\ifx.\@tempc\let\@tempd=\@citepunct\else
    \ifx,\@tempc\let\@tempd=\@citepunct\else
    \let\@tempd=\@tempa\fi\fi\@tempd}
\def\@citepunct{\@tempc\edef\@sf{\spacefactor=\the\spacefactor\relax}\@tempa
    \@sf\@gobble}

\def\citenum#1{{\def\@cite##1##2{##1}\cite{#1}}}
\def\citea#1{\@cite{#1}{}}

\newcount\@tempcntc
\def\@citex[#1]#2{\if@filesw\immediate\write\@auxout{\string\citation{#2}}\fi
  \@tempcnta\z@\@tempcntb\m@ne\def\@citea{}\@cite{\@for\@citeb:=#2\do
    {\@ifundefined
       {b@\@citeb}{\@citeo\@tempcntb\m@ne\@citea\def\@citea{,}{\bf ?}\@warning
       {Citation `\@citeb' on page \thepage \space undefined}}%
    {\setbox\z@\hbox{\global\@tempcntc0\csname b@\@citeb\endcsname\relax}%
     \ifnum\@tempcntc=\z@ \@citeo\@tempcntb\m@ne
       \@citea\def\@citea{,}\hbox{\csname b@\@citeb\endcsname}%
     \else
      \advance\@tempcntb\@ne
      \ifnum\@tempcntb=\@tempcntc
      \else\advance\@tempcntb\m@ne\@citeo
      \@tempcnta\@tempcntc\@tempcntb\@tempcntc\fi\fi}}\@citeo}{#1}}
\def\@citeo{\ifnum\@tempcnta>\@tempcntb\else\@citea\def\@citea{,}%
  \ifnum\@tempcnta=\@tempcntb\the\@tempcnta\else
   {\advance\@tempcnta\@ne\ifnum\@tempcnta=\@tempcntb \else \def\@citea{--}\fi
    \advance\@tempcnta\m@ne\the\@tempcnta\@citea\the\@tempcntb}\fi\fi}

\def\abstract{\if@twocolumn
\section*{Abstract}         
\else \small
\begin{center}
{ABSTRACT\vspace{-.5em}\vspace{0pt}}
\end{center}
\quotation
\fi}
\def\endabstract{\if@twocolumn\else\endquotation\fi}

\def\fnum@figure{Fig. \thefigure}

\long\def\@makecaption#1#2{
   \vskip 10pt
   \setbox\@tempboxa\hbox{\small #1. #2}
   \ifdim \wd\@tempboxa >\hsize    
      \small #1. #2\par            
   \else                           
      \hbox to\hsize{\hfil\box\@tempboxa\hfil}
   \fi}

\catcode`\@=12 

\def\defeq{\equiv}
\def\Toprel#1\over#2{\mathrel{\mathop{#2}\limits^{#1}}}
\def\Botrel#1\under#2{\mathrel{\mathop{#2}\limits_{#1}}}

\newcommand{\nubar}{\Toprel{\hbox{$\scriptscriptstyle(-)$}} \over{\nu}}

\newcommand{\qL}{q_{\scriptscriptstyle L}}
\newcommand{\qR}{q_{\scriptscriptstyle R}}
\newcommand{\qbL}{\bar{q}_{\scriptscriptstyle L}}
\newcommand{\qbR}{\bar{q}_{\scriptscriptstyle R}}

\newcommand{\eqref}[1]{(\ref{#1})}   

\newcommand{\GeV}{\hbox{$\hbox{GeV}^2$}}

\newcommand{\Ds}{\Delta s}

\newcommand{\syst}{~{\rm (syst.)}~}
\newcommand{\stat}{~{\rm (stat.)}~}

\def\beq{\begin{equation}}
\def\eeq{\end{equation}}
\def\bea{\begin{eqnarray}}
\def\eea{\end{eqnarray}}
\def\bq{\begin{quote}}
\def\eq{\end{quote}}

\def\gappeq{\mathrel{\rlap {\raise.5ex\hbox{$>$}}
{\lower.5ex\hbox{$\sim$}}}}

\def\lappeq{\mathrel{\rlap{\raise.5ex\hbox{$<$}}
{\lower.5ex\hbox{$\sim$}}}}

\begin{document}
\pagestyle{empty}
\begin{flushright}
CERN-TH.7022/93\\
TAUP 2094-93\\
hep-ph/9310272\\
\end{flushright}
\title{SPIN STRUCTURE FUNCTIONS
\thanks{Plenary talk at the
13th International Conference on Particles and Nuclei,
PANIC '93, 28~June -- 2~July 1993,  Perugia, Italy.}
}
\author{
JOHN ELLIS\\
{\em TH Division, CERN, Geneva, Switzerland \\
e-mail: johne@cernvm.cern.ch}
\\
\vspace{0.3cm}
and \\
\vspace{0.3cm} MAREK KARLINER\\
{\em
School of Physics and Astronomy
\\ Raymond and Beverly Sackler Faculty of Exact Sciences
\\ Tel-Aviv University, 69978 Tel-Aviv, Israel
\\ e-mail: marek@vm.tau.ac.il
}}
\maketitle

\abstract{
    We review the theory and phenomenology of deep inelastic
polarized lepton-nucleon scattering in the light of recent data
with a deuteron target from the SMC at CERN and a Helium 3 target
from the E142 experiment at SLAC. After including higher-order
perturbative QCD corrections,
mass corrections and updated estimates of higher-twist
effects, we find good agreement with the basic Bjorken sum rule,
and
extract a consistent set of values for the quark contributions to
the proton spin:
$$ \Delta\Sigma  \equiv
\Delta u + \Delta d + \Delta s = 0.27 \pm 0.11 $$
$$
\Delta u = 0.82 \pm 0.04~, ~~\Delta d = -0.44 \pm 0.04~,~~\Delta s =
-0.11 \pm 0.04
$$
which are consistent with chiral soliton models and indications from
lattice estimates. We also
mention the prospects for future experiments on the spin structure of
the nucleon. }
\vspace{0.5cm}
\begin{flushleft}
CERN-TH.7022/03\\
TAUP 2094-93\\
September 1993\\
\end{flushleft}
\vfill\eject
\setcounter{page}{1}
\pagestyle{plain}

\section{Introduction}
Polarized lepton-nucleon scattering is characterized by two
spin-dependent structure
functions $G_{1,2}$, as follows:
\beq
\frac{d^2\sigma^{\uparrow\downarrow}}{dQ^2d\nu} -
\frac{d^2\sigma^{\uparrow\uparrow}}
{dQ^2d\nu} = \frac{4\pi\alpha^2}{Q^2E^2}~\bigg[M_N(E+E^{\prime}\cos
\theta )G_1(\nu ,Q^2)
- Q^2G_2(\nu ,Q^2)\bigg]
\label{1}
\eeq
According to the na\"\i ve
parton model, these structure functions have simple scaling behaviours
in the Bjorken
scaling limit
\beq
x = \frac{Q^2}{2M_N\nu}~~{\rm fixed,}~~Q^2\rightarrow\infty
\label{2}
\eeq
given by
\bea
{\nu\over M^2_N} G_1(\nu ,Q^2) \defeq g_1(x,Q^2)
\rightarrow g_1(x) \nonumber\\
\label{3} \\
\left( { \nu\over M^2_N}\right)^2
G_2(\nu ,Q^2) \defeq g_2(x,Q^2) \rightarrow g_2(x)
\nonumber
\eea
These scaling structure functions can be related to the distributions of
quarks with
spins parallel and antiparallel to that of the target nucleon
\bea
g^p_1(x) &=& {1\over 2} \sum_q~e^2_q[q_{\uparrow}(x) - q_{\downarrow}(x)
+ \bar
q_{\uparrow}(x) - \bar q_{\downarrow}(x)]\\ \nonumber
&=&  {1\over 2} \sum_q~\Delta q(x)
\label{4}
\eea
For comparison, in the Bjorken scaling limit the unpolarized structure
function can be
written as
\beq
F_2(x) = \sum_q e^2_qx[q_{\uparrow}(x) + q_{\downarrow}(x) + \bar
q_{\uparrow}(x) -
\bar q_{\downarrow}(x)]
\label{5}
\eeq
Polarized lepton-nucleon scattering experiments actually measure the
polarization
asymmetry
\beq
A_1 = \frac{\sigma_{1/2}-\sigma_{3/2}}{\sigma_{1/2} + \sigma_{3/2}}
\label{6}
\eeq
where $\sigma_{3/2}$ and $\sigma_{1/2}$ are the cross-sections for
scattering with the
spin of the photon parallel and antiparallel to the spin of the
longitudinally-polarized
nucleon. In the Bjorken scaling limit, the polarization asymmetry in Eq.
(\ref{6}) can
be written as
\beq
A_1(x) =
\frac{\sum_q~e^2_q~[q_{\uparrow}(x) - q_{\downarrow}(x) + \bar
q_{\uparrow}(x) - \bar
q_{\downarrow}(x)]}
{\sum_q~e^2_q~[q_{\uparrow}(x) + q_{\downarrow}(x) + \bar
q_{\uparrow}(x) + \bar
q_{\downarrow}(x)]}
\label{7}
\eeq
Thus the asymmetry measurements in polarized lepton-nucleon scattering
experiments must be
combined with independent measurements of the unpolarized structure
functions in other
experiments in order to extract $g_{1,2}$\ , as we discuss later.

Much of the interest in deep inelastic polarized lepton-nucleon
scattering arises from
its relationship to axial current matrix elements. Neglecting for the
moment
perturbative QCD complications in the singlet axial current sector, one
can represent
the different quark axial currents matrix elements as follows:
\beq
\langle p\vert A^q_{\mu}\vert p\rangle = \langle p\vert\bar
q\gamma_{\mu}\gamma_5q\vert
p\rangle = \langle p\vert\qbR \gamma_{\mu}\qR -
\qbL\gamma_{\mu}\qL\vert
p\rangle = \Delta q\cdot S_{\mu}(p)
\label{8}
\eeq
where $q_{\scriptscriptstyle L,R}
\equiv 1/2 (1 \mp \gamma_5) q$, $S_{\mu}$ is the nucleon
spin
four-vector, and
\beq
\Delta q \equiv \int^1_0 dx[q_{\uparrow}(x) - q_{\downarrow}(x) + \bar
q_{\uparrow}(x) -
\bar q_{\downarrow}(x)]
\label{9}
\eeq
of particular interest is the flavour-singlet axial current
\beq
A^0_{\mu} = \sum_{q=u,d,s} \bar q\gamma_{\mu}\gamma_5q:\quad\quad
\langle p\vert A^0_{\mu}\vert p\rangle =
\sum_{q=u,d,s} \Delta q\cdot S_{\mu}(p)
\label{10}
\eeq
where $\Delta\Sigma\equiv
\lower-0.1em\hbox{$\mathop{\scriptscriptstyle\sum}\limits_q$}\Delta q$
is na\"\i vely
interpreted as the sum of the quark contributions to the proton spin.

It is worthwhile to recall here the general spin
decomposition of the proton
\beq
{1\over 2} = {1\over 2} \sum_q \Delta q  + \Delta G + \langle L_z\rangle
\label{11}
\eeq
where the three terms on the right-hand side represent the contributions
of quarks,
gluons and orbital angular momentum to the proton helicity in the
infinite momentum
frame, loosely referred to simply as the proton spin. We note in passing
that the
contribution of $L_z$ is not negligible in nuclei
such as the deuteron and $^3$He, as we shall discuss in more detail
later
on.

The first pieces of experimental
information about the $\Delta q$ came from charged-current weak
interactions. In particular, neutron $\beta$-decay, together with an
innocent $SU(2)$
isospin transformation, leads to\cite{RPP}
\beq
\Delta u - \Delta d = F+D = 1.2573 \pm 0.0028
\label{12}
\eeq
where $F$ and $D$ are the two independent irreducible
matrix elements of the axial currents in $SU(3)_f$.
Hyperon $\beta$-decays, together with a somewhat less innocent
$SU(3)$ flavour
transformation yield\cite{FoverD}
\beq
\frac{\Delta u + \Delta u - 2\Delta s}{\sqrt{3}}
= {3F -D \over \sqrt{3}} = 0.34 \pm 0.02
\label{13}
\eeq
We note in passing that the same hyperon $\beta$-decays yield
\beq
F/D = 0.58 \pm 0.02
\label{14}
\eeq
to be compared with the values of 2/3 in the na\"\i ve constituent quark
model and 5/9
in chiral soliton model.

As we shall see, equations (\ref{11}) and (\ref{12}) together
provide two equations
for the three unknowns $\Delta u, \Delta d$ and $\Delta s$. As we shall
discuss in more
detail later on, polarized lepton-nucleon scattering provides a third
equation which
enables $\Delta u, \Delta d$ and $\Delta s$ to be determined. However,
it is
interesting to observe\cite{EK,KaplanManohar}
that there is an alternative source of the third
equation.
Elastic $\nubar p \rightarrow \nubar p$ scattering depends on the $Z^0$
coupling to the
proton, which has a piece proportional to the axial current $\bar u
\gamma_u \gamma_5
u-\bar d\gamma_{\mu}\gamma_5 d -  \bar s\gamma_{\mu}\gamma_5 s$. Thus,
the axial $Z^0$
coupling in the $Q^2\rightarrow 0$ limit measures the combination
$\Delta u - \Delta d - \Delta s$.
The experimental data presently available\cite{Ahrens}
enable only a rough estimate to be made:
\beq
\Delta s = -0.15 \pm 0.09
\label{15}
\eeq
but a new, high-precision experiment is now being prepared, which should
enable
$\Delta s$ to be measured with an accuracy of
$\pm0.03\syst\pm0.03\stat$
\cite{GarveyPrivate}

Deep-inelastic polarized lepton-nucleon scattering should obey certain
sum rules, of
which the most basic is that derived by Bjorken\cite{BJ}
\beq
\int^1_0 dx \bigg[g^p_1 (x,Q^2) - g^n_1(x,Q^2)\bigg] = {1\over 6}
(\Delta u-\Delta
d)\times \bigg(1-{\alpha_s(Q^2)\over\pi} \bigg) + \ldots
\label{16}
\eeq
where we have indicated explicitly
the leading-order perturbative QCD correction and the dots
represent subasymptotic corrections,
which will be
discussed in more detail later. This sum rule, derived by Bjorken in the
late 1960's,
was the original motivation for scaling in the Bjorken limit. It is an
essential
prediction of QCD,\cite{Kodaira}
and all QCD theorists would have to eat their
collective hat it if
turned out to be violated.

It is possible to derive additional sum rules only by making further
dynamical
assumptions. Precisely because such additional assumptions are
necessary, the
theoretical foundation of such additional sum rules is much less firm
than that of the
Bjorken sum rule. Specifically, it was proposed in Ref.~[\citenum{EJ}]
that $\Delta s =
0$, in which case
\beq
\int_0^1 dx~g^p_1 (x,Q^2) = {1\over 18} (4\Delta u + \Delta d)~\bigg( 1
- {\alpha_s(Q^2)\over \pi}\bigg) + \ldots \quad = 0.17 \pm 0.01
\label{17}
\eeq
at $Q^2 = 10.7$ GeV$^2$.
The assumption of Ref.~[\citenum{EJ}] was based on intuition provided by
the
na\"\i ve
constituent quark model. It seemed reasonable at the time to assume that
there were no
strange quarks in the proton, and if there were, that surely they would
have no net
polarization. The original motivation for writing down the sum rule was
that a new
generation of experiments with polarized beam and target was about to
start at SLAC,
and it would be helpful to have some qualitative idea of
what could be seen in those experiments.
Nowadays, we are clear that this sum rule is
not a fundamental
test of QCD, but depends on an assumption about a non-perturbative
hadronic matrix
element, that could be and indeed seems to be wrong.

\section{Polarized Proton Data}

Pioneering experiments were carried out in 1976-1983 by a SLAC-Yale
collaboration,\cite{oldSLACa,oldSLACb,oldSLACc}
yielding the estimate
\beq
\int^1_0 g^p_1(x,Q^2) = 0.17 \pm 0.05
\label{18}
\eeq
This result was inconclusive as a test of the sum rule in
Eq.~(\ref{16}), because of the
large error bars, a large part of which was due to the extrapolation to
$x = 0$.  The
next round came in 1987, with the EMC result\cite{EMCPL,EMCNP}
\beq
\int^1_0 g^p_1(x,Q^2) = 0.126 \pm 0.010 \syst\pm 0.015\stat\ , \quad
\langle Q^2\rangle = 10.7 ~{\rm GeV}^2
\label{19}
\eeq
which is significantly different from the prediction of
Ref.~[\citenum{EJ}]
evaluated in \hbox{Eq.~(\ref{17})} with the updated values of $F$ and
$D$,
in (\ref{12}) and (\ref{13}).
This
experiment
actually measured polarized muon-proton scattering over the range
$0.01 < x < 0.7$,
extending down to lower values of $x$ than in the SLAC-Yale experiments.
The EMC data
showed a substantial deviation at low $x$ from the predictions of some
theoretical models
that had been assumed by the SLAC-Yale collaboration. However, the EMC
behaviour at
low $x$ is consistent with expectations based on Regge behaviour
\beq
g^p_1(x) \simeq \sum_i x^{-\alpha_i(0)}\beta^{\gamma}_i\beta^N_i
\label{20}
\eeq
where $\beta_i^{\gamma}$ and $\beta^N_i$ are the couplings to the photon
and nucleon of
the $i$-th Regge trajectory, and the $\alpha_i(0)$ are the corresponding
Regge
intercepts. A knowledge of meson spectra and exchanges leads us to
expect\cite{Heimann} $\alpha_i(0) \simeq 0$ to $-0.5$,
whilst a fit to the low-$x$ region of the EMC data
yields\cite{EK}
\beq
g^p_1 \sim x^{-\delta} : \quad \delta = -0.07^{+0.42}_{-0.32}~~{\rm
for}~~x < 0.2
\label{21}
\eeq
This consistency gives us no reason to doubt the EMC value of the sum
rule shown in Eq.~(\ref{19}).
One would expect similar low-$x$ behaviour in the
neutron structure
function.

The EMC polarized muon-proton data provide a third equation
\beq
{1\over 2} \bigg({4\over 9} \Delta u + {1\over 9} \Delta d + {1\over
9}\Delta
s\bigg)\,\bigg(1 - {\alpha_s(Q^2)\over \pi} \bigg) + \ldots = 0.126 \pm
0.010\syst\pm 0.015\stat
\label{22}
\eeq
with which we can determine\cite{EFR,EMCPL}
the three quark contributions to the proton
spin
\bea
\Delta u &=& \phantom{-}0.78 \pm 0.06 \nonumber \\
\Delta d &=& -0.47 \pm 0.06 \label{23} \\
\Delta s &=& -0.19 \pm 0.06 \nonumber
\eea
We see that $\Delta s \not= 0$, and the sum rule of Ref.~[\citenum{EJ}]
is
clearly violated.
Adding together the different contributions in Eq. (\ref{23}),
we find that the total contribution of quarks to the proton spin is
\beq
\Delta\Sigma =
\Delta u + \Delta d + \Delta s = 0.12 \pm 0.17
\label{24}
\eeq
which is consistent with zero. This has sometimes been referred to in
the literature as
the ``spin crisis". This is an exaggeration. The result in Eq.
(\ref{24}) is
certainly a surprise for our original na\"\i ve understanding
of non-perturbative QCD, but does not conflict with any rigorous result
of perturbative
QCD.

\section{Theoretical Interpretation}

By now we are familiar with competing models of hadronic structure: the
na\"\i ve
constituent quark model on the one hand, and the chiral soliton models
on the other
hand. Certain aspects of baryon  and nuclear phenomenology are
better
described by the former, and others by the latter class of models.
Neither holds a
monopoly of truth.

In the na\"\i ve non-relativistic quark model (NRQM) one thinks of the
proton  or
neutron as a composite of three relatively heavy, slow-moving
constituent quarks, with
$m_{p,n} \simeq 3m_q$ and $m_q \simeq 300$ MeV. In particular, the spin
of the proton
or neutron is obtained by combining na\"\i vely the spins of the three
non-relativistic
constituent quarks, which
can be depicted schematically as
\beq
p,n^{\Uparrow} = q^{\Uparrow}q^{\Uparrow}q^{\Downarrow}
\label{25}
\eeq
The NRQM yields good values for the anomalous magnetic moments of the
proton and
neutron, and has been very successful in describing hadron spectroscopy.
However, the
model can be justified rigorously in QCD only for very heavy quarks such
as the $b$
or $t$.

Other aspects of hadron physics, in particular the low-energy
interactions of pions,
are well described by approximate chiral symmetry, which would become
exact if the
quark masses in the underlying QCD Lagrangian were to vanish. The small
physical value
of the pion mass is related to the smallness of the $u$ and $d$ quark
masses
\beq
m_\pi^2 = \langle 0 | \bar{q} q | 0 \rangle
\,{m_u + m_d\over f_\pi^2}
\label{26}
\eeq
The chiral symmetry picture can be extended to nucleons with the aid of
the $1/N_c$
expansion\cite{tH,WittenB},
an expansion in the inverse of the number of colours in QCD.
This
combination justifies a view of the nucleon as a soliton ``lump" of
light
pseudoscalar meson fields\cite{Skyrme,SM,WittenLewes}
-- a ``skyrmion".
This model gives good or acceptable values
for the ratios of
proton and neutron magnetic moments\cite{SM} and
successful predictions\cite{ourPiN,SiegenPiN}
of meson-nucleon scattering phase shifts.
In the Skyrme model, the proton wave function may be viewed as
\beq
\vert p\rangle \simeq V(t)U(r)V^{-1}(t)~;\quad U = \exp [iF(r)\hat
r\cdot r]
\label{27}
\eeq
where $V(t)$ is a time-dependent $SU(3)$ flavour matrix which
represents a slow
collective rotation in the flavour space. As it is based on an effective
chiral
Lagrangian, it is expected to be good for reproducing the ``soft"
(low momentum transfer) properties of nucleons, such as axial current
matrix elements.
According to this picture, the proton contains many relativistic quarks,
and the
angular momentum of the nucleon
is due to the slow collective rotation of the soliton, parametrized by
$V(t)$ in (\ref{27}).
As was pointed out in Ref.~[\citenum{BEK}],
it is a general feature of such chiral soliton models that
the nucleon matrix elements of the flavour-singlet axial current Eq.
(\ref{8}) are
identically zero, implying that the net contribution of quarks to the
proton or nucleon spin
vanishes,
\beq
\langle N\vert A^0_{\mu}\vert N\rangle = 0
\quad \Longrightarrow \quad \Delta\Sigma
= \sum_q \Delta q = 0
\label{28}
\eeq
This result follows directly from the topology of the flavour group
manifold, and has
nothing to do with the perturbative $U(1)$ axial anomaly of Eq. (30)
below. In the Skyrme
model and its simple extensions there are no gluons,
and therefore $\Delta G = 0$ identically. Thus the angular momentum sum
rule of Eq.
(\ref{17}) becomes\cite{EK}
\beq
{1\over 2} = {1\over 2}  \Delta\Sigma  (=0) + \Delta G(=0) + \langle
L_z\rangle~(={1\over 2})
\label{29}
\eeq
in such chiral soliton models. Of course, orbital angular momentum can
only take
integer values in any given parton state, so that the statement $L_z =
1/2$ refers to the
expectation value of the orbital angular momentum carried by the quarks,
after appropriate
statistical weighting of all the states contributing to the proton wave
function. The fact
that $L_z$ is not an integer in such a strongly-coupled system should
not come as a
surprise. After all, even in the weakly-bound deuteron there is a 5 \%
admixture of the
$D$-wave in the nuclear wave function. The prediction Eq. (\ref{29}) is
valid in the
large-$N_c$ limit and for massless quarks. We expect in general
corrections coming from
$1/N_c$ terms and from finite quark masses, especially the strange quark
mass, $m_s \simeq$
150 MeV $\simeq \Lambda_{QCD}$. Estimates
indicate\cite{BEK,Ryzak,SkyrmeNow}
that these could modify the prediction of
Eq. (\ref{29}) by $\lappeq$ 30 \%.

We would like to emphasize that the
above-mentioned
soliton picture of the nucleon is not necessarily in conflict with the
NRQM. The fact that
\hbox{$\Delta s \not= 0$}
experimentally (\ref{23}) and Eq. (\ref{24}) could be
reconciled with the
NRQM if constituent quarks have internal
structure\cite{GM,Kaplan,Fritzsch,EFHK,staticQ}.
It is possible to model this structure
in a type of chiral model where the chiral field carries both flavour
and colour\cite{Kaplan,EFHK,staticQ}. The
constituent quarks then emerge as solitons in such a chiral model, just
as the nucleon
emerges as a soliton in the usual chiral Lagrangian. It is also possible
to model the chiral
constituent quarks in the Nambu-Jona-Lasinio model, with similar
results\cite{Weise}.

Shortly after the EMC results were published, an important theoretical
observation was made.\cite{DeltagI,DeltagII,DeltagIII}
Consider the flavour-singlet axial current $A^0_{\mu}$. This
current is conserved
at the classical level if one neglects the small current quark masses,
just like the
flavour non-singlet currents. However, it has a non-zero divergence at
the quantum
level, due to the one-loop triangle anomaly
\beq
\partial^{\mu} A^0_{\mu} = {\alpha_s\over \pi}~N_f~{\rm tr}~
F_{\mu\nu}\tilde F^{\mu\nu}~,
\quad \tilde F^{\mu\nu} \equiv
\epsilon^{\mu\nu\alpha\beta}F_{\alpha\beta}
\label{30}
\eeq
where $F_{\mu\nu}$ is the gauge field strength tensor in QCD. The
intuitive meaning of
Eq. (\ref{30}) is that the anomaly induces a mixing between gluons and
the
flavour-singlet axial current of quarks. For this reason, the helicities
carried by
each flavour $\Delta u, \Delta d$ and $\Delta s$ undergo  additive
renormalization,
\bea
\Delta u \longrightarrow \widetilde{\Delta u} = \Delta u -
(\alpha_s/2\pi )\Delta G
\\ \nonumber
\Delta d \longrightarrow \widetilde{\Delta d} = \Delta d -
(\alpha_s/2\pi )\Delta G
\\ \nonumber
\Delta s \longrightarrow \widetilde{\Delta s} = \Delta s -
(\alpha_s/2\pi )\Delta G
\label{31}
\eea
Despite appearances, these corrections are {\it not} suppressed by a
power of
$\alpha_s$. The reason is that $\Delta G \sim \log Q^2$. Whereas
$\alpha_s$ decreases
as $Q^2$ increases, $\Delta G$ increases, and the product $\Delta\Gamma
\equiv
(\alpha_s/2\pi )\Delta G$ is $Q^2$-independent in leading order. Recall
also that, since
the $\Delta G$ term is the same for all flavours, it does not contribute
to flavour
non-singlet matrix elements, for example $\widetilde{\Delta u} -
\widetilde{\Delta d} =
\Delta u - \Delta d$.

Some physicists have proposed\cite{DeltagII}
that this effect could rescue the NRQM
intuition, by making
$\Delta s = 0$ compatible with experiment. This is possible in
principle, because
experiment only tells us the value of $\widetilde{\Delta s}$. If $\Delta
G$ were large
enough, one could have $\Delta s = 0$, with the observed non-zero value
of
$\widetilde{\Delta s}$ induced by the gluon term alone. However, this
would require
$\Delta G \simeq 4$ (in units of $\hbar$)
at $Q^2 =$ 10 GeV$^2$, which seems rather unlikely.
The angular
momentum sum rule of Eq. (\ref{17}) would then read
\beq
{1\over 2} = {1\over 2} \sum_q \Delta q + \Delta G (\simeq 4) + \langle
L_z\rangle~~
(\simeq -4)
\label{32}
\eeq
which is rather counter-intuitive. On top of this, the anomaly
interpretation has to
face an additional difficulty. A detailed calculation of the relevant
box diagram shows\cite{Box}
that the contribution of $\Delta G$ to the first moments of $\Delta
q(x)$ tends to
concentrate at low values of $x$.
Since the EMC
experiment covers only a limited range of $x$, the contribution of gluon
polarization
to the {\it observed}~$\Delta q$ is actually much smaller than suggested
by Eq. (\ref{31}),
and hence an even larger value of $\Delta G$ is required.

We stress again that the chiral soliton model and the anomaly
interpretation of the EMC
data are two physically distinct possibilities.
The former is based on the topology of the $SU(3)$ flavour
group, whereas the latter is based on quantum breakdown of
the classical $U_A(1)$ symmetry.
It is possible in
principle to
distinguish between them by measuring the gluon polarization in the
nucleon. The
required experiments are difficult, but are within the reach of current
experimental
techniques.

\section{Analysis of New Data\protect{\cite{EKBjSR}} }

When attempting to check experimentally a QCD prediction for a sum rule,
one has to
deal with the following problem, due to a very basic mismatch between
the
theory and
what is actually measured. Theoretical predictions for sum rules are
always formulated
at a fixed value of $Q^2$. A generic sum rule in QCD typically reads
\beq
\Gamma (Q^2) = \Gamma_{\infty} \bigg[ 1 + \sum_{n \ge 1}
c_n~\bigg({\alpha_s(Q^2)\over\pi}\bigg)^n\bigg] + \sum_{m \geq
1}~{d_m\over (Q^2)^m}
\label{33}
\eeq
where $\Gamma_{\infty}$ is the asymptotic value of the sum rule for
$Q^2\rightarrow\infty$, the $c_n$ are the coefficients of the
perturbative corrections,
and the $d_m$ are  coefficients of the so-called mass and higher-twist
corrections.
On the other hand, in a typical experiment, data are taken at {\it
variable} values of
$Q^2$, with a significant and monotonic correlation between the range of
$Q^2$ and the
value of $x$, as shown schematically in Fig.~1. This results from the
finite
kinematical range of any given experiment. Moreover, as has already been
mentioned,
polarized lepton-nucleon scattering experiments measure directly the
polarization
asymmetry defined in Eq. (\ref{6}), rather than the polarized structure
functions shown in
Eq.~(\ref{1})$\,$.

To overcome this limitation and correct simultaneously  for the
{above-}\break mentioned
$x$-dependence of the range of $Q^2$,
it is convenient to use the experimental fact\cite{EMCPL,EMCNP}
that $A_1(x,Q^2)$ seems
approximately independent of $Q^2$, and reconstruct the
polarized structure
function $g_1(x,Q^2)$ using
\beq
g_1(x,Q^2) = {A_1(x,Q^2)F_2(x,Q^2)\over 2x [1 + R(x,Q^2)]}
\simeq
{A_1(x)F_2(x,Q^2)\over 2x [1 + R(x,Q^2)]}
\label{34}
\eeq
where $R(x, Q^2)$
is the ratio of longitudinal to transverse virtual photon
cross-sections. One can take $F_2(x,Q^2)$ and $R(x,Q^2)$ from previous
high-precision parametrizations of unpolarized data. Equation
(\ref{34}) and the assumption of $Q^2$-independence of $A_1$ imply that
the $Q^2$ dependence of $g_1(x,Q^2)$ is determined by the
$Q^2$-dependences of $F_2(x,Q^2)$ and of $R(x,Q^2)$. In particular, the
higher-twist effects well known to occur in $F_2$ and $R$ at low $Q^2$
will show up in $g_1$. Clearly, the approximation of Eq. (\ref{34}) is
most reliable at $Q^2 = \langle Q^2\rangle$, but we will also need to
make the same assumption at other values of $Q^2$, in order to combine
the proton and neutron data, which have been taken at different values
of $Q^2$.

The procedure of Eq. (\ref{34}) has been used previously by the EMC to
interpret their polarized $\mu$-$p$ data. Now  one must reevaluate those
asymmetry data, incorporating more recent parametrizations of
$F_2(x,Q^2)$
by the NMC\cite{NMC},
and of $R(x,Q^2)$ from SLAC\cite{RLT}. We use these to evaluate
\beq
\Gamma_1^p(Q^2) \equiv \int^1_0 dx~g^p_1(x,Q^2)
\label{35}
\eeq
at different values of $Q^2$ as seen below
\bea
Q^2 ({\rm GeV}^2) & \Gamma_1^p(Q^2) &          \nonumber \\
                               \nonumber \\
       \matrix{2.0   \cr 4.6   \cr 10.7  }
       \phantom{123}
&\left.\matrix{0.124 \cr 0.125 \cr  0.128}
\right\}
&\pm 0.013 \pm 0.019
\label{36}
\eea
The first two values of $Q^2$ are chosen for comparison with the recent
SMC data\cite{SMC} at $\langle Q^2 \rangle = 4.6$ GeV$^2$
and E142 data\cite{E142} at $\langle Q^2 \rangle = 2$ GeV$^2$.
The last value of $Q^2$ is that used
previously by the EMC\cite{EMCPL,EMCNP},
and our value of $\Gamma_1^p(Q^2 = 10.7$
GeV$^2$) is well within the errors they originally quoted.

We apply the same procedure to re-interpret the E142  polarized
$e$-$^3$He data\cite{E142},
interpreted as $e$-$n$ asymmetry data after the nuclear
structure
corrections discussed below.  When we rescale to fixed $Q^2 = 2 $
GeV$^2$, using the NMC\cite{NMC} and
SLAC\cite{RLT} parametrizations mentioned earlier, we find
\beq
\int^{0.6}_{0.03} dx~g^n_1(x,Q^2{=}2~{\rm GeV}^2) =
-0.022\pm 0.006 \stat \pm 0.006\syst
\label{37}
\eeq
whose central value is again very close to that quoted by the  E142
collaboration.

No experiment can measure the full range $0 \leq  x \leq 1$, and every
experiment must make some assumption in order to extrapolate to the
full $x$ range. Non-perturbative models of neutron structure suggest
that $A_1^n(x)\rightarrow 1$ as $x\rightarrow 1$, but there is no
indication of this from perturbative QCD. Therefore, we prefer to be
agnostic and allow the asymmetry to vary within the possible kinematic
range $\vert A_1^n(x)\vert   \leq 1$. We then find a high-$x$
contribution to $\Gamma_1^n$ for the E142 experiment of
\beq
\Delta\Gamma_1^n = 0.000 \pm 0.003
\label{38}
\eeq
The extrapolation of the E142 data to $x = 0$ is {\it a priori} more
uncertain than that of the SMC, because of a larger lower limit on $x$
(0.03 to be compared with 0.006). Motivated by the polarized proton
data discussed in Section 2, we assume the following plausible Regge
form for the low-$x$ extrapolation
\beq
g^n_1(x) = Ax^{\alpha}~:~~0 \leq \alpha \leq 0.5
\label{39}
\eeq
which yields a low-$x$ contribution to the E142 integral of
\beq
\Delta\Gamma_1^n = -0.006 \pm 0.006
\label{40}
\eeq
Putting together Eqs. (\ref{37}), (\ref{38}) and (\ref{40}), we arrive
at the
following final estimate of the polarized neutron integral at $Q^2 = 2$
GeV$^2$
\beq
\Gamma_1^n(Q^2{=}2~{\rm GeV}^2) =
-0.028 \pm0.006 \stat \pm 0.009\syst
\label{41}
\eeq
which is consistent within errors with the value quoted by the E142
collaboration.

In the case of the SMC experiment, which used the procedure (\ref{34})
with up-to-date
parametrizations of $F_2(x,Q^2)$ and $R(x,Q^2)$, we find
\beq
\Gamma_1^n(Q^2{=}4.6~{\rm GeV}^2) =
-0.076 \pm 0.046\syst \pm 0.037\stat
\label{42}
\eeq
The only difference from their published paper is due to the
re-evaluation in Eq. (\ref{36}) of the original EMC proton data, that
are used
in a subtraction.

Before using the above numbers, some comments are in order on the
corrections due to the nuclear structure of $^3$He and deuterium.
Na\"\i vely, one would view the $^3$He nucleus as containing a pair of
protons with paired spins, and an odd neutron which carries all the
nuclear spin. However, a general description of the
$^3$He nuclear wave
function contains 10 components. This can be simplified for most
practical purposes to a three-component description with a $D$-wave
component that varies between 8.6 and 9.8 \% (a typical estimate
and comprehensive discussion can be found in
Ref.~[\citenum{Ciofi}])
an $S^{\prime}$ component that varies in strength between
1.4 and 1.7 \%, and the rest made up by the conventional $S$-wave
component. The above numbers lead to the following plausible estimates
for the mean polarizations of protons and neutrons in the $^3$He nucleus
in the Bjorken
scaling limit
\beq
p_p = (-2.5 \pm 0.3) \%~, ~~ p_n = (87 \pm 2) \%
\label{43}
\eeq
This then yields the following relation between the integrals of the
polarized structure functions for neutrons, $^3$He and protons
\beq
\Gamma^n = (1.15 \pm 0.02)\Gamma^3 + (0.057 \pm 0.009)\Gamma^p
\label{44}
\eeq
Calculations indicate that this nuclear
structure correction is almost independent of $x$ in the Bjorken
scaling limit, and the above estimates indicate that uncertainties in
this nuclear structure correction are not important in the analysis of
the E142 data. However, one caveat should be mentioned. The above
estimates are in the Bjorken scaling limit, and there may well be large
effects at finite $Q^2$, which could lead to significant corrections to
the E142 data.\cite{CiofiPrivate}
Similar effects are also possible in principle for the
deuteron,\cite{Tokarev}
but should be less important, as it is a simpler nucleus and
the SMC data are at larger values of $Q^2$.

We conclude this section by mentioning an interesting bound on
$g_1^n(x,Q^2)$, derived in Ref.~[\citenum{PRbound}].
The idea of this bound is to
eliminate $\Delta u(x,Q^2)$ between
$g_1^p(x,Q^2)$ and $g_1^n(x,Q^2)$. Neglecting higher-twist effects, one
finds
\beq
\bigg\vert 4g^n_1(x,Q^2)-g^p_1(x,Q^2)\bigg\vert{=}\bigg\vert {15\over
18}~
\Delta d(x,Q^2) + {3\over 18}~ \Delta s(x,Q^2)\bigg\vert{\leq}{15\over
18}
d(x,Q^2) + {3\over 18} s(x,Q^2)
\label{45}
\eeq
When using this equation, one cannot simply assume that the deuteron
structure function is the sum of proton and neutron structure
functions, but must include the deuteron nuclear structure correction
\beq
\Gamma_1^p(Q^2) + \Gamma_1^n(Q^2) \simeq {\Gamma_1^d(Q^2)\over 1-1.5
\omega_D}
\label{46}
\eeq
where $\omega^D$ is the probability of finding the deuteron in a
$D$-wave, $\omega^D \simeq 0.058$, and one must also bear in mind the
existence of smearing at large $x$.  We choose to express Eq. (\ref{45})
in
terms of the following directly measured quantities: $A_1^p(x)$,
$A_1^n(x)$, $R(x,Q^2)$, and $\xi(x,Q^2) \equiv
F_2^N(x,Q^2)/F_2^p(x,Q^2)$
\beq
\bigg\vert4A^n_1(x,Q^2) \xi (x,Q^2) - A^p_1(x,Q^2)\bigg\vert
\leq \bigg[ 1+R(x,Q^2)\bigg]~\bigg[4\xi (x,Q^2)-1\bigg]
\label{47}
\eeq
To check the inequality, one should evaluate all the terms using the
same
fixed value of $Q^2$. As seen in Fig.~2a, we find that the SLAC E142
data are highly consistent with this inequality at $Q^2 = 2$ GeV$^2$,
whereas Fig.~2b indicates that
there is a very marginal disagreement
($\ll 1 \sigma$) between the SMC and EMC
data at $Q^2 = 4.6$ GeV$^2$. We emphasize that this
discrepancy cannot be considered significant, in view of the large
experimental errors and the possible existence of higher-twist
and mass corrections.

\section{More Theoretical Interpretation\protect{\cite{EKBjSR}}}

Combining the EMC and E142 data we find
\beq
\Gamma_1^{p-n}(Q^2 = 2~{\rm GeV}^2) = 0.152 \pm 0.014 \pm 0.021
\label{48}
\eeq
whilst combining the EMC and SMC data we find
\beq
\Gamma_1^{p-n}(Q^2 = 4.6~{\rm GeV}^2) = 0.201 \pm 0.048 \pm 0.042
\label{49}
\eeq
These evaluations at fixed $Q^2$ of the Bjorken sum rule integral must
be compared with the theoretical prediction
\bea
\Gamma_1^{p-n}(Q^2) = && {1\over 6} \bigg\{ g_A
\bigg[1-{\alpha_s(Q^2)\over\pi}
- 3.58 \bigg({\alpha_s(Q^2)\over\pi}\bigg)^2 - 20.2
\bigg({\alpha_s(Q^2)\over\pi}\bigg)^3+\ldots \bigg] \nonumber\\
&& \phantom{123}
- {8\over 9Q^2} \bigg( \ll U^{NS} \gg + {1\over2} m_N^2
\ll V \gg  \bigg)
\bigg\}  \label{50}\\
&& + {4\over 9} ~{m^2_N\over Q^2}~\int^1_0 dx~x^2 g_1^{p-n} (x,Q^2)
+\ldots
\nonumber
\eea
where we have included the full set of perturbative QCD corrections
calculated in Ref.~[\citenum{BJcorr}],
as well as the leading higher-twist term and the
leading target-mass corrections.\cite{HigherTwist,BBK,Ji,IanPrivate}
The higher-twist term is obtained from
\beq
\langle N\vert U^{NS}_{\mu}\vert N \rangle \equiv S_{\mu} \ll U^{NS} \gg
\label{51}
\eeq
where
\beq
U^{NS}_{\mu} \equiv \bar ug\tilde G^a_{\mu\nu}\gamma_{\nu}~{1\over 2}
\lambda^a
u - (u\rightarrow d)~:~~\tilde G^a_{\mu\nu}\equiv
\epsilon_{\mu\nu\alpha\beta}G^{\alpha\beta}_a
\label{52}
\eeq
and an analogous expression for $\ll V \gg$\cite{BBK}.
The first estimate of the matrix elements $\ll U^{NS} \gg$
and $\ll V^{NS} \gg$ was given in Ref.~[\citenum{BBK}].
This initial estimate has been criticized as either too
small\cite{BI} or too large\cite{Ji}. Following this criticism,
the calculation in Ref.~[\citenum{BBK}] has been recently re-checked
by one of the authors and the most recent
estimate\cite{IanPrivate} is
\beq
- {8\over 9Q^2} \bigg( \ll U^{NS} \gg + {1\over2} m_N^2
\ll V \gg  \bigg)
\simeq - {0.1\over Q^2}
\label{53}
\eeq
to which we assign the same error as was quoted\cite{BBK}
previously, namely \break
$\pm 0.15\,\hbox{GeV}^2/Q^2$.
In the following we will use the value given in Eq.~(\ref{53}),
while keeping in mind the possibility
that it might change in either direction.

Eq.~(\ref{53}) makes a significant
correction to the Bjorken sum rule at $Q^2 = 2$ GeV$^2$, but is
negligible compared with the SMC errors at  $Q^2 = 4.6$ GeV$^2$. The
integral in the target mass correction in Eq. (\ref{51}) is evaluated to
be
0.0168 at $Q^2 = 2$ GeV$^2$ and 0.0130 at $Q^2 = 4.6$ GeV$^2$.
Using a recent determination of $\alpha_s(Q^2)$ from $\tau$
decays\cite{alphasRef} and
incorporating all the finite-$Q^2$ corrections in Eq. (\ref{50}), we
find the
following theoretical predictions
$$
\Gamma_1^{p-n}(Q^2 = ~~2~{\rm GeV}^2) = 0.160 \pm 0.014
\eqno{(54a)}
$$
$$
\Gamma_1^{p-n}(Q^2 = 4.6~{\rm GeV}^2) = 0.177 \pm 0.007
\eqno{(54b)}
$$
as seen in Fig.~3.
Confronting these with the experimental values shown in Eqs. (\ref{48})
and
(\ref{49}), we conclude that {\it the Bjorken sum rule is satisfied
within one standard deviation}.

The perturbative QCD and higher-twist machinery used above can be
cross-checked\cite{CK}
with the Gross-Llewellyn Smith sum rule. The CCFR
collaboration has recently published a new evaluation\cite{CCFR}
of this sum rule
at $Q^2 = 3$ GeV$^2$
\addtocounter{equation}{1}
\beq
\int^1_0 dx~F_3^{\bar\nu p+\nu p}(x,Q^2 = 3~{\rm GeV}^2) = 2.50 \pm
0.018 \pm
0.026
\label{55}
\eeq
to be compared with the na\"\i ve prediction of 3 in the
quark-parton model. Including perturbative QCD corrections\cite{BJcorr}
and higher-twist effects\cite{BK}, one finds\cite{CK}
for the right-hand-side of Eq.~(\ref{55})
\beq
3\bigg[ 1 - {\alpha_s(Q^2)\over\pi} -
3.58\bigg({\alpha_s(Q^2)\over\pi}\bigg)^2
- 19.0  \bigg({\alpha_s(Q^2)\over\pi}\bigg)^3 - {8\over 27} ~~{\ll
O^s\gg\over Q^2} + \ldots \bigg]
\label{56}
\eeq
where the higher-twist coefficient is given by
\beq
\langle P\vert O^s_{\mu}\vert P\rangle \equiv 2p_{\mu} \ll O^s \gg
\label{57}
\eeq
where
\beq
O^s_{\mu} = \bar u~\tilde G_{\mu\nu} \gamma_{\nu}\gamma_su +
\bar d~\tilde G_{\mu\nu} \gamma_{\nu}\gamma_sd
\label{58}
\eeq
The higher-twist coefficient has been estimated\cite{BK} as
\beq
\ll O^s \gg = 0.33~{\rm GeV}^2
\label{59}
\eeq
with a precision of perhaps 50\%. Figure~4 shows that there is
very good agreement between experiment, Eq. (\ref{55}), and theory, Eq.
(\ref{56}), and it is claimed that this agreement may even be improved
by
including the higher-twist effect. These data have in fact been used to
determine\cite{CK}
\beq
\alpha_s(m_z) = 0.115 \pm 0.006
\label{60}
\eeq
in the $\overline{\rm MS}$ renormalization scheme, in good agreement
with other determinations from deep-inelastic scattering, LEP and
elsewhere.

One way of stating the agreement of the polarized structure function
data with the Bjorken sum rule is to extract an effective value of
$g_A$. This is done by subtracting the higher-twist and mass
corrections from the data, and then removing the perturbative QCD
correction factors. Following this procedure, we find from the EMC and
SMC data
\beq
g_A = 1.43 \pm 0.45
\label{61}
\eeq
and from the EMC and SLAC E142 data
\beq
g_A = 1.20 \pm 0.22
\label{62}
\eeq
Combining all three experiments we find
\beq
g_A = 1.24 \pm 0.20
\label{63}
\eeq
and conclude that {\it the Bjorken sum rule is verified to within 16
\%}.

Given this high degree of consistency with the fundamental Bjorken sum
rule, we now proceed to extract the different quark contributions to
the nucleon spin. We do this by using the EMC proton data, the SMC
proton plus neutron data after making the deuteron $D$-wave correction,
and the E142 data for the neutron after  making the $^3$He wave
function correction discussed earlier.
The higher-order perturbative corrections
to the singlet integral $\Gamma^{p+n}(Q^2)$ are not available, but the mass
corrections and leading higher-twist effect have been calculated. The latest
updated estimate\cite{IanPrivate} of the latter is
$\simeq 0.1\,\hbox{GeV}^2/Q^2$ with a large error, which we take here to
be the same as in Ref.~[\citenum{BBK}].

As in the extraction of the
effective value of $g_A$, we first subtract from the data the
theoretical values of the 1/$Q^2$ corrections, and then remove the
perturbative QCD factors.
We denote the resulting moments with a tilde, $\tilde\Gamma_1$,
to distinguish them from the raw experimental quantities
$\Gamma_1$.
Following this procedure, we find
$$
\phantom{12}
\tilde\Gamma_1^p(Q^2 = 10.7~{\rm GeV}^2) = {1\over 2}~({4\over
9}~\Delta u +
{1\over 9}~\Delta d + {1\over 9}~\Delta s) = \phantom{{-}}
0.138 \pm 0.023
\eqno{(64a)}
$$
$$
\phantom{2}
\tilde\Gamma_1^n(Q^2 = \phantom{1}2.0~{\rm GeV}^2)~
= {1\over 2}~({1\over 9}~\Delta
u +
{4\over 9}~\Delta d + {1\over 9}~\Delta s) ={-}0.045 \pm 0.016
\eqno{(64b)}
$$
$$
\tilde\Gamma_1^{n+p}(Q^2 = \phantom{1}4.6~{\rm GeV}^2) = {1\over
2}~({5\over
9}~\Delta u
+ {5\over 9}~\Delta d + {2\over 9}~\Delta s) =\phantom{{-}}
 0.051 \pm 0.055
\eqno{(64c)}
$$
 Combining these
results with neutron-$\beta$ decay
\addtocounter{equation}{1}
\beq
g_A = \Delta u - \Delta d = 1.2573 \pm 0.0028
\label{65}
\eeq
and hyperon beta-$\beta$ assuming $SU(3)$
as discussed in Section 2, we find three independent estimates of the
sum of the quark contributions to the proton spin
$$
\Delta\Sigma  \equiv
\Delta u + \Delta d + \Delta s =
\left\{
\matrix{
0.15 \pm 0.21,\qquad\qquad \hbox{(EMC, $Q^2=10.7$ \GeV)}
\hskip1.0em(66a)\cr
\cr
0.08 \pm 0.25,\qquad\qquad \hbox{(SMC, $Q^2=
\phantom{1}4.6$ \GeV)} \hskip1.0em(66b)\cr
\cr
0.39 \pm 0.14,\qquad\qquad \hbox{(EMC, $Q^2=
\phantom{1}2.0$ \GeV)} \hskip1.0em(66c)}
\right.
$$
The different determinations of $\Delta\Sigma$ and
$\Delta s$ are plotted in Fig.~5, where we see a high degree
of consistency.
The world average of $\Delta\Sigma$ is
\addtocounter{equation}{1}
\beq
\Delta\Sigma = 0.27 \pm 0.11
\label{67}
\eeq
with individual contributions of
\beq
\Delta u = 0.82 \pm 0.04~, ~~\Delta d = -0.44 \pm 0.04~,~~\Delta s =
-0.11 \pm 0.04
\label{68}
\eeq
We see that the total quark contribution to the proton spin
is positive but small, and that the
strange quark contribution is significantly non-zero.

In this context it is interesting to note that $\Ds$ being nonzero is
a part of an intriguing pattern \cite{IK}:
experiment indicates that certain strange-quark bilinear operators,
such as $\bar s \gamma_\mu\gamma_5 s$ have relatively large
matrix elements in the proton, while others are very small.
The presence of a substantial non-valence component of
$\bar{s} s$ pairs in the proton has some striking consequences.
One of these is the evasion of the OZI rule in the couplings of
$\bar{s} s$ mesons to baryons \cite{EGK}, leading to
surprisingly large branching ratios for $\phi$ production in
$\bar {p} p$ annihilation at rest \cite{pbarpI}.

At this point we note that there are three possible attitudes
towards three higher-twist effects: one is simply to ignore them, like
the ostrich. Another is to treat the coefficient of the higher-twist
correction as a free parameter\cite{Close}, setting its
value through the requirement
that the EMC, SMC and E142 results for $\Delta \Sigma$
are consistent with each other,
in which case the available data yield
\beq
\Delta\Sigma = \Delta u + \Delta d + \Delta s = 0.38 \pm 0.48
\label{69}
\eeq
We have taken a third approach, which is to use the best available
theoretical calculations to produce Eq.~(\ref{68}).

\section{Conclusions and Prospects}

A wealth of new data on polarized lepton-nucleon structure functions
are now being accumulated. All the data available so far are consistent
with QCD, and the Bjorken sum rule is verified with a precision of
about 16 \%. There is good convergence between the different
measurements of the quark contributions to the proton spin, which are
summarized in Eqs. (\ref{67}) and (\ref{68}). We note that the value of
$\Delta\Sigma$ is qualitatively consistent with the predictions of
chiral soliton models\cite{BEK}
and lattice simulations\cite{lattice0,latticeI,latticeII,latticeIII}.
The most recent lattice calculation\cite{latticeIII} finds
$\Delta\Sigma = 0.10 \pm 0.21$.

We welcome the more precise data on polarized lepton-proton and
-neutron structure functions that will become available shortly,
including data on the interesting structure function $g_2$. We
emphasize that cleaner tests of QCD are possible at higher $Q^2$,
corresponding to higher beam energies, since the higher-twist and mass
corrections vanish asymptotically. Therefore, other things being equal,
preference should be given to running with higher-energy beams.

Also of interest for the future are data on polarization asymmetries
for final-state particles. Measuring $\pi^+/\pi^-$ asymmetries should
confirm the values of $\Delta u$ and $\Delta d$ discussed above,
whereas measuring $K^+/K^-$ asymmetries should probe $\Delta s$
directly.
Particularly interesting would be polarization asymmetries
for charmed final states, including the $J/\psi$, which would probe
$\Delta G$ directly.

Forthcoming deep inelastic experiments include a continuation of the
SMC experiment using protons in 1993 and other targets in subsequent
years, the SLAC E142 experiment taking data on proton and deuteron
targets in 1993, and the SLAC E143 experiment taking data on proton and
deuteron targets with a 50 GeV beam in 1995.\cite{E143}
A particularly interesting
newcomer will be the Hermes experiment at HERA\cite{HERMES}
using proton, deuteron
and $^3$He targets from 1995 onwards. This experiment will have
particular advantages for studying final-state asymmetries.
In the longer
term, there is a proposal\cite{LEP} to perform polarized lepton-nucleon
scattering experiments in LEP. Finally, we mention the ambitious
project for injecting polarized protons into RHIC and colliding beams
each
of 250 GeV. Measuring jet and Drell-Yan asymmetries one should be
able to measure $\Delta q$ and $\Delta G$ directly.\cite{CollinsPanic}

Polarized lepton-nucleon scattering experiments have already provided
us with plenty of interest and some surprises. We are sure that this
field will continue to excite the community of physicists for the
foreseeable future.
\bigskip
\begin{flushleft}
{\bf Acknowledgements}
\end{flushleft}
The  research of M.K. was supported in part
by grant No.~90-00342 from the United States-Israel
Binational Science Foundation (BSF), Jerusalem, Israel,
and by the Basic Research Foundation administered by the
Israel Academy of Sciences and Humanities.
\bigskip

\def\etal{{\em et al.}}
\def\PL{{\em Phys. Lett.\ }}
\def\NP{{\em Nucl. Phys.\ }}
\def\PR{{\em Phys. Rev.\ }}
\def\PRL{{\em Phys. Rev. Lett.\ }}

\noindent
FIGURE CAPTIONS

\begin{description}

\item [Fig. 1] In any given polarized lepton-nucleon scattering
experiment,
the range of $Q^2$ probed is different in different bins of the Bjorken
variable $x_{B_j}$.

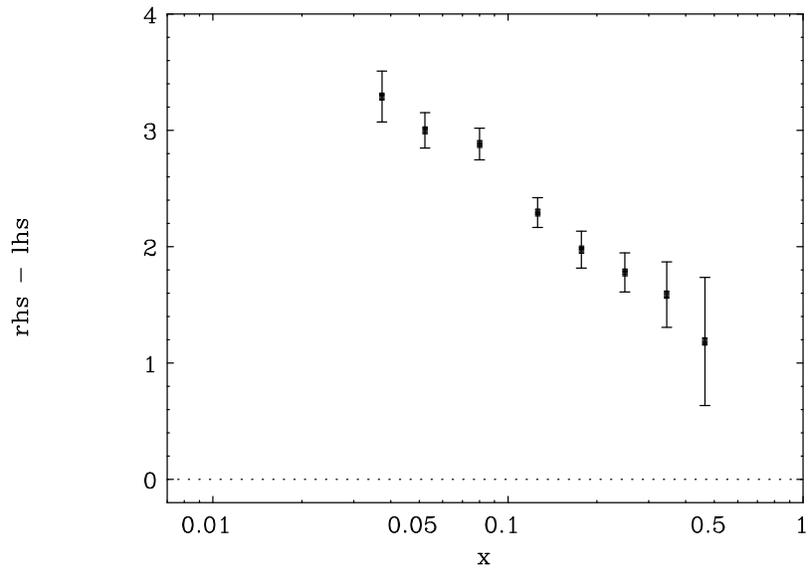
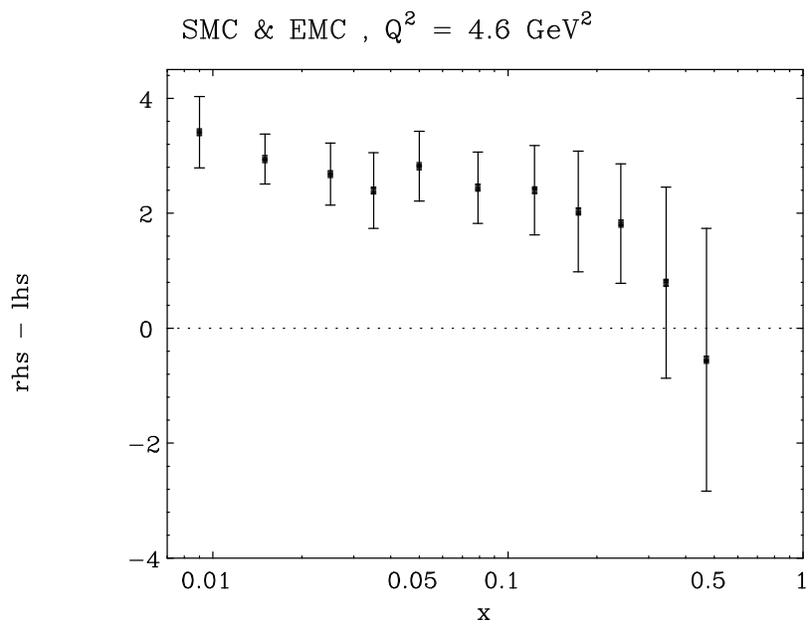
\item[Fig. 2] The difference between the right-hand and left-hand sides
of the bound \cite{HigherTwist}.  The actual errors are slightly larger
than those indicated
by the error bars, as the latter refer to the error in the left-hand
side only.  (a) E142 data
\cite{E142}, combined with EMC data \cite{EMCPL}, rescaled to $Q^2 =
2$~GeV$^2$;  (b) SMC data
\cite{SMC}, combined with EMC data \cite{EMCPL}, rescaled to $Q^2 =
4.6$~GeV$^2$.

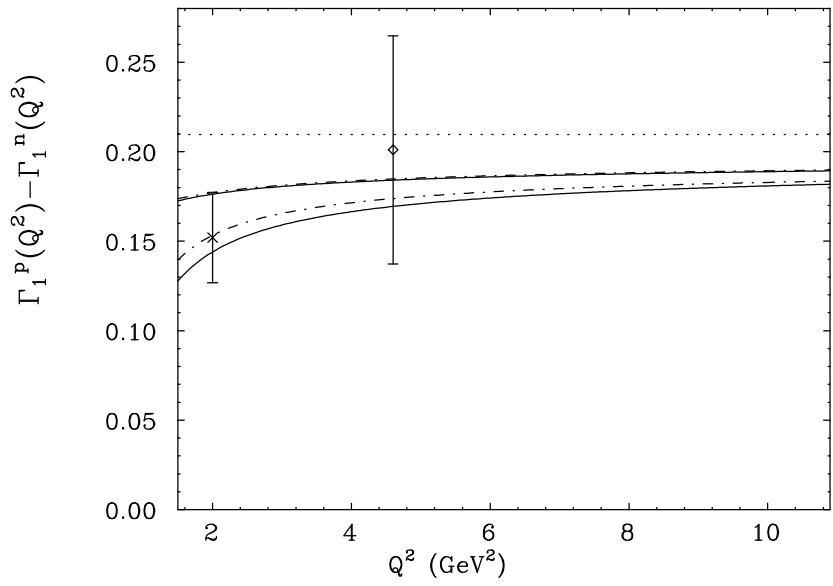
\item[Fig. 3] Experimental tests at $Q^2$ = 2 GeV$^2$:  E142 and EMC,
$Q^2$
= 4.6~GeV$^2$:  SMC and EMC, of the Bjorken sum rule, including
perturbative
QCD corrections (dot-dashed lines) and higher-twist corrections (solid
lines).  The asymptotic value $g_A/6$ is denoted by a dotted line.

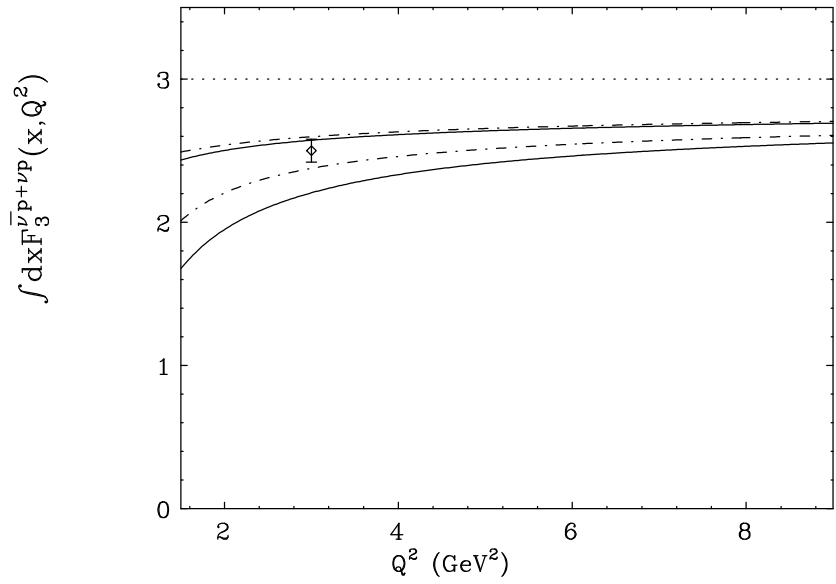
\item[Fig. 4] Experimental test \cite{CCFR} at $Q^2$ = 3~GeV$^2$ of the
Gross-Llewellyn Smith
sum rule including perturbative QCD corrections (dot-dashed lines) and
higher-twist corrections
(solid lines).

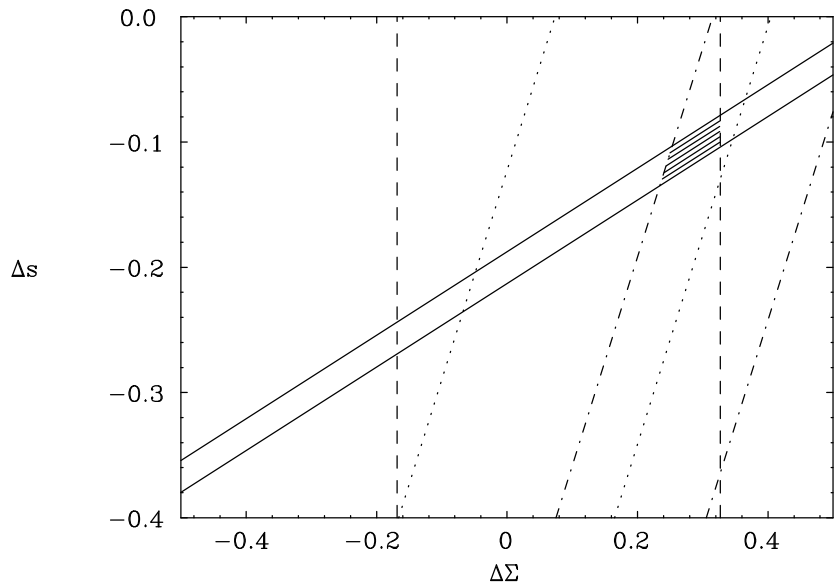
\item[Fig. 5] The allowed regions in the $\Delta\Sigma - \Delta s$
plane, corresponding to the
linear constraints (64) and (\ref{13}).  Continuous lines:
$\Delta \Sigma - 3 \Delta
s = 3F - D$;  dots:  $\tilde \Gamma^p_l(Q^2)$ constraint;  dot-dash:
$\Gamma^n_l(q^2)$;  dashes:  $\tilde \Gamma^p_l(q^2) + \tilde
\Gamma^n_l(q^2)$.

\end{description}

\end{document}